# SocIoS API: A data aggregator for accessing user generated content from online social networks

Magdalini Kardara[1], Vasilis Kalogirou[1], Athanasios Papaoikonomou[1], Theodora Varvarigou[1], Konstantinos Tserpes[2]

[1]National Technical University of Athens, Dept of Electrical and Computer Engineering, Athens, Greece
`{nkardara, vaskalogirou, tpap, dora}@mail.ntua.gr`
[2]Harokopio University of Athens, Dept of Informatics and Telematics, Greece
`tserpes@hua.gr`

**Abstract.** Following the boost in popularity of online social networks, both enterprises and researchers looked for ways to access the social dynamics information and user generated content residing in these spaces. This endeavor, however, presented several challenges caused by the heterogeneity of data and the lack of a common way to access them. The SocIoS framework tries to address these challenges by providing tools that operate on top of multiple popular social networks allowing uniform access to their data. It provides a single access point for aggregating data and functionality from the networks, as well as a set of analytical tools for exploiting them. In this paper we present the SocIoS API, an abstraction layer on top of the social networks exposing operations that encapsulate the functionality of their APIs. Currently, the component provides support for seven social networks and is flexible enough to allow for the seamless addition of more.

## 1   Introduction

Online communities, such as social networking and social media platforms, have experienced an outstanding boost in their popularity which in turn resulted to the existence of a large amount of online content created by the members of such communities. This content does not only refer to data, such as text posts, photos and videos, which are explicitly shared by the users online. It also involves a significant amount of social information related to users and implicitly derived by their actions, such as their interests and preferences as well as their relationships with other users.

Although some of this content remains private, a significant amount of it is made publicly available by its owners. More importantly, the social networks themselves, instead of limiting access and usage of their functionality and data, they have propagated them freely, making them part of their core offering to end users and at the same time allowing third parties to build applications on top of them. Currently the most popular social networks and media, such as Twitter [1], Facebook [2] and YouTube [3] expose all or part of their functionality through open RESTful APIs through which every user or third party application can gain access to their content and operations.

Both enterprises and researchers have long recognized the huge potential of the social graph information and user created content residing in social networks and looked for ways to harness them. For enterprises with appropriate tools for managing them, the data available in social network are potential sources of revenue, as they can be valuable assets in targeted advertising and viral marketing campaigns. In research, the popularity of social networks has brought new interest in various research domains, as the vast amount of user-generated content and the explicit connections between users allows the study of data analysis and social dynamics on an unprecedented scale.

Although the social networking content is ample and easily accessible, harnessing this content still presents several challenges. Despite the similarities in notions and basic functionality, data representation in social networks is highly heterogeneous. In addition to that, each social network offers its own API and due to the lack of a non commercial tool for accessing multiple APIs from a single API, a user looking to combine data from two or more social networks will have to invoke all the APIs and transform the data in a common format before processing them.

The SocIoS framework aims to address the abovementioned challenges. It is a software stack that operates on top of Social Networking Sites (SNS). SocIoS provides an abstraction layer for combining data and functionality from a multitude of underlying social media platforms as well as a set of analytical tools for leveraging that functionality.

At the core of the SocIoS project, lies the SocIoS API. It constitutes a single access point for a number of popular social networks exposing operations that encapsulate their functionality. For each supported SNS a respective adaptor has been developed. Currently, the component provides support for seven social networks: Facebook, Twitter, FlickR, Dailymotion, YouTube, Google+ and Instagram. Support for additional social networks can be added by implementing the adaptor interface provided.

The remainder of this paper is structured as follows: Section 2 presents related work in the field of social network interoperability; Section 3 gives an overview of the SocIoS Framework; Section 4 describes the SocIoS object model; Section 5 focuses on the internal design of the SocioS API; Section 6 presents conclusions and future work.

## 2 Related Work

The diversity of SNS APIs and data object models necessitate a meta-API that will act as an aggregation point and provide seamless access to the whole spectrum of User Generated Content (UGC). In the market, there are commercial products that fulfill this need. First, GNIP [4] a company that was acquired by Twitter in April 2014, provides access to numerous social data sources, both in real-time and historical mode. Similar approaches are followed by HootSuit [5] and DataSift [6] which enable their users to manage multiple social networks and also offer various analytics services. In academia, a number of research projects have contributed tools with analogous capabilities, like +Spaces [7] [8] and the toolbox [9] developed by WeGov [10] which exploit social networking technologies for policy making, as well as SOCIETIES [11] which facilitates the creation of user communities with integrated social networking capability [12].

A popular approach when it comes to aggregating profiles in SNS, are the aggregator websites: one-stop shopping sites for SNSs that provide users with a common interface for accessing multiple social networks [13]. The main differentiation of these websites from API aggregators is the fact that they do not seek for a semantic aggregation of the main concepts in the SNSs but instead they wrap the APIs' outcome in a single interface without further analyzing them. Instead, the SocIoS platform, approaches the problem from an ontological point of view, by first attempting to identify the common notions in the underlying SNSs [14] and then wrapping the API calls in a single object model.

A relevant work towards the creation of a single SNS ontology has been conducted by Mika [15] in 2005. In Mika's work the starting point are not the existing SNSs but instead a theoretic structure of a social network. The adaptation of this to popular social media platforms should require significant effort since their concepts have been developed independently. In our approach we preferred the use of an object model deriving from the analysis of the operating online social networks. The OpenSocial specification [16] inspired the creation of a new object model which allowed the easy wrapping of underlying API calls to a new, single meta-API call. This approach is better analyzed in what follows.

## 3   SocIoS Framework and Approach

The SocIoS framework is a software stack that operates on top of SNSs with the purpose of:

- Aggregating data and functionality from a multitude of underlying social media platforms,
- Providing a tool for developers to build social analytics services on top of the supporting social media platforms,
- Accommodating newly created applications that use the abovementioned services and provide them through usable interfaces.

With the proper configuration and development of intermediate services, the framework can support any application that requires the harvesting of social media, filtering the content using sophisticated features while at the same time harnessing the scale issues of the endeavor (volume of data, number of users and platforms).

The objectives mentioned above are achieved through a layered Service-Oriented Architecture (SOA) which is depicted in Fig. 1. The SocIoS Framework consists of several main entities, such as the SocIoS API, the Auxiliary Services(i.e. Data Analy-sis and Added Value Services) and the Front-End. A short description of these components is given below.

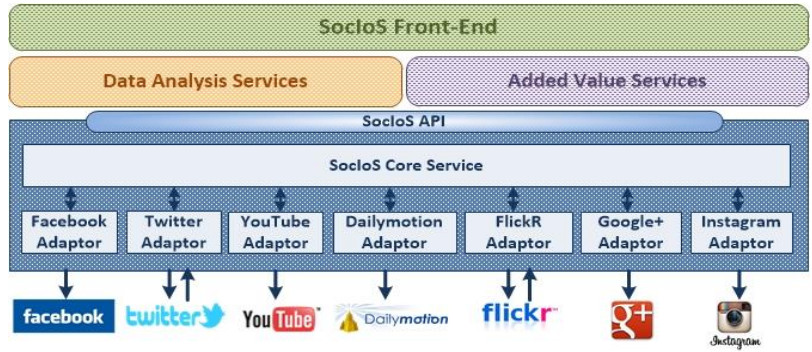

**Fig. 1.** SocIoS platform high level architecture

An important artifact of the SocIoS Framework, not depicted above, is the SocIoS Object model, an abstraction of the data models defined by the underlying social network APIs. This is used by all the SocIoS components for representing entities and the relations between them.

The SocIoS API is the core part of the framework, exposing a set of operations the higher service layers (SocIoS Auxiliary Services) and mapping them to collections of methods and objects of the social networking site APIs. It consists of a set of adaptors that perform the data transformation between the supported SNSs' APIs and the Object Model and a central component, the Core Service, responsible for coordinating the adaptors.

The Auxiliary Services are third party services that extend the functionality of the core services and use the same data models as the core services. Depending on their purpose, their operations may or may not be exposed as part of the SocIoS API in the sense that they can extend it. This differentiates them between integrated and non-integrated Auxiliary Services. The integrated services are data analysis services, providing analytics to the data delivered by the Core Services, while the non-integrated ones are added value applications that enhance the business potential of SocIoS. The Auxiliary Services are standalone components that can be re-used, they were, however, developed with the purpose to meet the requirements of a certain application. Some examples are:

- Media Item Ranking and Recommendation (Data Analysis): A service for assessing the subjective value of a media item to a specific user [17],
- Topic-Specific Community Detection (Data Analysis): A service for identifying social media user communities that are implicitly linked to each other [18],
- Event Detection (Data Analysis): A service for highlighting intense or unusual activity within a community [19],
- Social Filtering (Data Analysis): A service for managing social media user groups [20],

- The FlexiPrice service (Added Value): A service that enables two users (a buyer and a seller) to set a price of a content item [21],
- The Crowdsourcing Game (Added Value): A service that allows the posting of crowdsourcing tasks in the form of a game [22].

Finally, the SocIoS Front-End provides a user interface for interacting with the components as well as an authentication mechanism. It also deals with the user rights and privacy settings while in certain applications, it may also serve as a point of integration for the Core and Auxiliary Services.

In what follows, we emphasize on the definition of the SocIoS API, i.e. the part that is implemented by the core services.

## 4  SocIoS Object Model

Each social network has its own data model for representing entities (users, posts etc) and the relationships between them. By examining the data models of some popular social networks, we come across objects that describe similar notions. For example, a photo uploaded in a social network and a video uploaded in another, can both be considered as "media items" and they both display similar properties, such as title, upload time, number of supportive manifestations etc. By identifying all the conceptually common objects, we can define a new data model and use it to capture data coming from any of the underlying social networks.

To this end, we introduce the Socios Object Model, a set of entities representing principal common notions in social networks. The Object Model consists of the following objects:

- **Person**: The Person object represents a user's profile on a social network. It contains the user's account and profile information.
- **MediaItem**: The MediaItem object represents a post published by a user in a social network. It can refer to a video, image or text post.
- **Activity**: The Activity object contains information about an action performed by a user in a social network.
- **Comment**: The Comment object represents a comment on a media item published in a social network.
- **SocialNetwork**: The SocialNetwork is an enumeration of the supported social networks, namely FLICKR, FACEBOOK, TWITTER, YOUTUBE, DAILYMOTION, GOOGLEP and INSTAGRAM
- **ObjectId**: The ObjectId is a wrapper that defines an object in a specific social network.
- **Address**: The Address object contains details about a location.
- **Name**: The Name object contains information about the user's full name.
- **License**: The License object contains basic information about the license attached to a media item.

| SocIoS Main Objects | | | | |
|---|---|---|---|---|
| Person | MediaItem | Activity | Comment | SocialNetwork |
| id | id | id | id | FLICKR |
| sn | sn | sn | sn | FACEBOOK |
| aboutMe | created | created | created | TWITTER |
| addresses | title | title | description | YOUTUBE |
| birthday | thumbnailUrl | description | userId | DAILYMOTION |
| currentLocation | description | location | username | GOOGLEP |
| username | duration | actorId | numPositiveVotes | INSTAGRAM |
| email | location | objectType | | |
| gender | language | mediaItems | | |
| name | license | persons | | |
| photos | fileSize | activities | | |
| profileUrl | rating | | | |
| memberSince | numRatings | | | |
| thumbnailUrl | numPositiveVotes | | | |
| utcOffset | numNegativeVotes | | | |
| numFriends | numComments | | | |
| inDegree | numViews | | | |
| outDegree | numResharings | | | |
| | numFavorites | | | |
| | tags | | | |
| | taggedPeople | | | |
| | type | | | |
| | url | | | |
| | userId | | | |
| | comments | | | |

**Fig. 2:** SocIoS Object Model - main objects

| SocIoS Secondary Objects | | | |
|---|---|---|---|
| ObjectId | Address | Name | License |
| id | country | firstName | licenseType |
| socialNetwork | extendedAddress | lastName | name |
| | latitude | additionalName | url |
| | longitude | fullName | |
| | postalCode | | |
| | region | | |
| | streetAddress | | |

**Fig. 3:** SocIoS Object Model - secondary objects

**Fig. 2** and **Fig. 3** display the abovementioned objects and their fields. Each object defined in the model maps to one or more entities in each social network, while some objects are only present in a subset of the underlying networks (for example only Google+, Facebook, YouTube and Dailymotion have the equivalent of an Activity).

As can be seen, the objects can be conceptually divided to main and secondary, with the former representing core social network concepts and the latter typically corresponding to complex fields belonging to the main objects. In addition to those, we have defined several filter objects, each one containing a set of conceptually relat-ed parameters that can be used for searches. These objects are typically used as input parameters in search related methods. The Filter Objects are depicted in Fig. 4.

| SocIoS Filters | | | | | | |
|---|---|---|---|---|---|---|
| PersonFilter | MediaItemFilter | ActivityFilter | AreaFilter | AddressFilter | LocationFilter | DateTimeFilter |
| keywords sns | created keywords location language licenseType sns | keywords language sns | latitude longitude radius | country postalCode region | addressFilter locationFilter | from to |

**Fig. 4:** SocIoS Object Model – filters

## 5  SocIoS API

As mentioned above, the SocIoS API lies on top of several popular social networks and provides uniform access to their APIs. In terms of implementation, the SocIoS API is comprised of a central component acting as the single point of reference for the layers above, available both in SOAP and REST, and a set of adaptors. The adaptors essentially act as wrappers on top of individual SNS APIs with which they communicate through REST.

As expected, the component is largely dependent on the APIs of the underlying social networks. This refers both to the functionality provided by each social network, as well as to their performance and limitations (e.g. number of calls per time frame). The component retrieves all information in real time, without storing or caching any user related data. Where necessary, authentication parameters are passed to the component by the layers above, as explained later on in the paper.

The following sections give an overview of the SocIoS Object model and an insight into the architecture design of the SocIoS API.

### 5.1  API Methods

As explained before, the SocIos API exposes a number of methods for interacting with the APIs of the underlying social networks. Some of these methods attempt to get data regarding specific objects whereas others retrieve objects that match certain criteria. Therefore, the input parameters either determine an identification of the objects we try to fetch, or act as filters for the search. The returned object contains the results of the queries that succeeded and detailed exceptions the ones that failed (if any). There is also a method for posting data.

The implementation of each method for each social network is subject to the limitations imposed by the SocIoS APIs themselves. For example Activity related methods are only implemented by social networks supporting the notion of an activity (i.e. Google+, Facebook, YouTube and Dailymotion, as stated above). Moreover, the functionality of each method may differ considerably from social network to social net-work based on the functionality provided by the respective network. For example, depending on the specific implementation of each network, the list of persons returned by the findPersonsByMediaItem method, may include just the owner of the media item or a full list of the people who have commented/liked/been tagged on it.

```
SOAP API
GetPersons (List<ObjectId> personIds)
ConnectedPersons (ObjectId personId)
MyConnectedPersons (ObjectId personId)
FindPersons (PersonFilter personFilter, ObjectId mediaItemId, ObjectId activityId, ObjectId username)
GetMediaItems (List<ObjectId> mediaItemIds)
GetMediaItemsForUser (ObjectId personId, ObjectId username)
GetMediaItemsForPage (ObjectId pageId)
FindMediaItems (MediaItemFilter mediaFilter)
FindRelevantMediaItems (ObjectId mediaItemId)
GetActivities (List<ObjectId> activityIds)
GetActivitiesForUser (ObjectId personId)
FindActivities (ActivityFilter activityFilter)
GetComments (List<ObjectId> commentIds)
GetCommentsForMediaItem (ObjectId mediaItemId)
GetCommentsForActivity (ObjectId activityId)
PostMessage (ObjectId personId, String postText)
```

**Fig. 5:** SOAP API

```
REST API
getPerson - input params: {id, sn, format}
connectedPersons - input params: {id, sn, format}
myConnectedPersons - input params: {id, sn, format}
findPersonsByKeyword - input params: {keywords, sns, format}
findPersonsByUsername - input params: {username, sn, format}
findPersonsByMediaItem - input params: {id, sn, format}
findPersonsByActivity - input params: {id, sn, format}
getMediaItem - input params: {id, sn, format}
getMediaItemsForUser - input params: {id, sn, username, format}
getMediaItemsForPage - input params: {id, sn, format}
findMediaItems - input params: {from, to, keywords, country, lat, lon, rad, lang, lic, sns, format}
findRelevantMediaItems - input params: {id, sn, format}
getActivity - input params: {id, sn, format}
getActivitiesForUser - input params: {id, sn, format}
findActivities - input params: {keywords, lang, sns, format}
getComment - input params: {id, sn, format}
getCommentsForMediaItem - input params: {id, sn, format}
getCommentsForActivity - input params: {id, sn, format}
postMessage - input params: {id, sn, msg, format}
```

**Fig. 6:** REST API

In order to avoid the need for the users to authenticate themselves for every single call, most of the SocIoS methods only deal with public information and do not require authentication. The only two exceptions are the myConnectedPersons and the postMessage methods, where user authentication is required by the underlying APIs. For these two methods, it is necessary to provide a valid authentication token for the user retrieved through the use of the OAuth protocol. Essentially what this means, is that the user allows SocIoS to gain access to their personal data or perform some action on behalf of them. The procedure for retrieving the OAuth tokens for each network is implemented by the SocIoS Front End, where the tokens for registered users are stored and kept updated.

Fig. 5 and Fig. 6 present the SOAP and REST interfaces of the component respectively.

### 5.2 API Architecture

In terms of architecture, the SocIoS API consists of one central component, the Core Service, which acts as the single point of reference for the layers above, and a set of adaptors. The Core Service is responsible for instantiating and coordinating the adaptors, as well as for gathering the results and returning them back to the client. The adaptors essentially act as wrappers on top of individual social network APIs with which they communicate through REST. Adaptors are part of the API (rather than deployed as standalone service) and are therefore Java objects instantiated and configured by the core services for each method invocation.

Each method implemented by the core service and exposed through the API maps to one or a combination of simpler methods implemented by the adaptors, with each adaptor providing an SNS specific implementation of each method. All adaptors (existing and new ones) must implement the ISnsAdaptor interface which defines all the methods that an adaptor must implement. Currently the SocIoS API contains adaptors for seven popular social networks, i.e. Twitter, Facebook, FlickR, Dailymotion, YouTube, Google+ and Instagram. New adaptors can be easily added to the platform by creating a new implementation of the ISnsAdaptor interface and adding the name of the adaptor to the list of known social networks.

The SNS adaptors are instantiated and coordinated by the Core Services module. The Core Service is responsible for processing the parameter list and creating the input parameters that will be subsequently passed to each adaptor. For example, the GetPersons method gets as input a list of ObjectId type parameters, each of which contains an id and a social network, and returns a list of Person object.

Upon receiving the request, the Core Service will process the list of ObjectId's received with the request and split it to a number of smaller lists, equal to the distinct number of social networks contained in the initial list. The list that will be passed to each adaptor will only contain the IDs belonging to the respective social network. After preparing the input parameters, the Core Service will instantiate the respective list of adaptors and invoke them in order to gather the results and get them back to the user. This process is depicted in Fig. 7.

The results returned by the adaptors are combined into a single return object, which also contains the errors that occurred during the execution and returned back to the end user. This means that errors occurring to one adaptor, will not significantly affect the functionality of the rest of the adaptors.

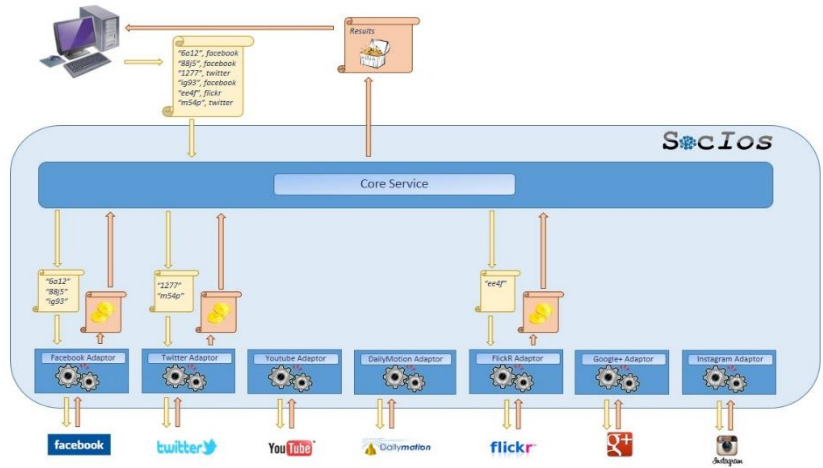

**Fig. 7:** Core Service and Adaptors

## 6   Conclusions and future work

In this paper we have presented SocIoS, a set of tools for aggregating and processing data from social networks. We described the SocIoS Object Model, a data model for representing social network entities and operations. We explained the functionality and implementation of the core component of the framework, i.e. the SocIoS API an abstraction layer on top of the social networks offering uniform access to their APIs. Currently the component supports seven popular social networks which can be easily extended to support more APIs with similar functionalities.

In the future we plan to extend the object model and API accordingly in order to accommodate more platforms that do not fall strictly under the social networking and media category, such as researchers communities and location based communities. The SocIoS API is maintained as a live project in github, from where it can be downloaded [20].

## 7   Acknowledgements

This work has been supported by the RADICAL project (http://www.radical-project.eu) and partly funded by the European Union's Competitiveness and Innovation Framework Programme under grant agreement no 325138.